\documentclass{article}
\usepackage[pdftex]{graphicx}
\usepackage[utf8]{inputenc}
\begin{document}

\title{Calculating the Hubble diagram by perturbation theory}
\author{Gyula Bene\\
Institute of Physics, Eötvös University\\
Pázmány P. s. 1/A, H-1117 Budapest, Hungary
}
\maketitle
\begin{abstract}
The effect of density fluctuations upon light propagation is calculated
perturbatively in a matter dominated irrotational universe. The starting point
is the perturbed metric (second order in the perturbation strength), while the
output is the Hubble diagram. Density fluctuations cause this diagram to
broaden to a strip. Moreover, the shift of the diagram mimics accelerated expansion.  
\end{abstract}
\section{Conventions}
We use the Landau conventions, i.e., we assume $+---$ signature for the metric. 
The zeroth component of a four-vector is timelike, 
the first, second and third components are spacelike. Four-vectors are indexed by Latin letters,
three-vectors by Greek letters. Throughout we use $c=1$ units.
\section{The Sachs optical equations}
A light ray may be parametrized along its path by a parameter $\lambda$ so
that its four-velocity $u^i$ is defined by
\begin{eqnarray}
u^i=\frac{dx^i}{d\lambda}\;.\label{e1.1}
\end{eqnarray}
It is a null vector,
\begin{eqnarray}
u^iu_i=0\label{e1.2}
\end{eqnarray}
and satisfies the geodesic equation
\begin{eqnarray}
u^i_{;k}u^k=0\;.\label{e1.3}
\end{eqnarray}
Let us introduce another independent null vector $w^i$ along the path. 
It satisfies
\begin{eqnarray}
w^iw_i=0\;,\label{e1.4}
\end{eqnarray}
\begin{eqnarray}
w^iu_i=1\label{e1.5}
\end{eqnarray}
and
\begin{eqnarray}
w^i_{;k}u^k=0\;.\label{e1.6}
\end{eqnarray}
Furthermore, let us define two spacelike unit vectors $L^i_\alpha$ ($\alpha,
\beta=1,2$) 
that are orthogonal to both the null
vectors $u^i$ and $w^i$:
\begin{eqnarray}
L^i_\alpha u_i=0\label{e1.7}
\end{eqnarray}
\begin{eqnarray}
L^i_\alpha w_i=0\label{e1.8}
\end{eqnarray}
\begin{eqnarray}
L^i_\alpha L_{i\beta}=-\delta_{\alpha \beta}\label{e1.9}
\end{eqnarray}
We also require that the vectors $L^i_\alpha$ are parallel translated along
the light ray: 
\begin{eqnarray}
L^i_{\alpha;k}u^k=0\label{e1.10}
\end{eqnarray}
Eqs. (\ref{e1.3}), (\ref{e1.6}), (\ref{e1.10}) ensure that if
Eqs. (\ref{e1.4}), (\ref{e1.5}), (\ref{e1.7}), (\ref{e1.8}), (\ref{e1.9}) are
satisfied at one single point of the light ray, they will be satisfied at any
other point of it as well.
For the separation $\xi^i$ of two light rays we assume
\begin{eqnarray}
\xi^iu_i=0\label{e1.11}
\end{eqnarray}
and
\begin{eqnarray}
\xi^iw_i=0\;,\label{e1.12}
\end{eqnarray}
hence we have
\begin{eqnarray}
\xi^i=\sum_{\alpha=1,2}d_\alpha L^i_\alpha\;.\label{e1.13}
\end{eqnarray}
The coefficients $d_\alpha$ describe the proper separation of the two
nearby light rays. For their derivative with respect to the path parameter
$\lambda$ we have in the most general case (since no rotation is possible)
\begin{eqnarray}
\frac{d d_\alpha}{d\lambda} = \sum \left(\Theta \delta_{\alpha \beta}+w_{\alpha \beta}\right)d_\beta\label{e1.14}
\end{eqnarray}  
where $w_{\alpha \beta}$ is symmetric and traceless:
\begin{eqnarray}
w_{\alpha \beta}=\left(\begin{array}{cc}\rho&\sigma \\ \sigma&-\rho\end{array}\right)\;.\label{e1.15}
\end{eqnarray}
Now, the geodesic deviation equation
\begin{eqnarray}
\frac{D^2\xi^i}{D\lambda^2}=R^i_{jkl}u^iu^k\xi^l\label{e1.16}
\end{eqnarray}
leads to the Sachs optical equations for the expansion rate $\Theta$ and shear
$w_{\alpha \beta}$:
\begin{eqnarray}
\frac{d\Theta}{d\lambda}+\Theta^2+\frac{1}{2}w^2=-\frac{1}{2}R_{jk}u^ju^k\label{e1.17}
\end{eqnarray}
\begin{eqnarray}
\frac{dw_{\alpha\beta}}{d\lambda}+2\Theta w_{\alpha\beta} =C_{ ijkl}L^i_\alpha
u^ju^kL^l_\beta\label{e1.18}
\end{eqnarray}
where
\begin{eqnarray}
w^2=w_{\alpha\beta}w_{\beta\alpha}\label{e1.19}
\end{eqnarray} 
and $C_{ ijkl}$ stands for the Weyl tensor.

The physical situation (e.g., that the two nearby light rays are
emitted from a point source) is specified by the initial conditions. After having solved these
equations, $\Theta$ governs the change of the cross section $A$ of a light
beam:
\begin{eqnarray}
\frac{dA}{d\lambda}=2\Theta A\;.\label{e1.20}
\end{eqnarray}
\section{Perturbation expansion}
We are going to consider a narrow light beam that propagates in a spacetime
with a slightly perturbed FRW metric. Hence, the expansion is done over the
corrections of the metric, the zeroth order being the (flat) FRW metric. 
We expand all the above quantities as follows (if the zeroth order is
automatically zero, we leave that out):
\begin{eqnarray}
g_{ij}=g_{ij}^{(0)}+g_{ij}^{(1)}+g_{ij}^{(2)}\label{e2.1}
\end{eqnarray}
Explicitly, according to Ref. \cite{Russ:1996}
\begin{eqnarray}
g_{00}^{(0)}&=&1\;,\;g_{0\alpha}^{(0)}=0\;,\;g_{\alpha\beta}^{(0)}=-a^2(t)\delta_{\alpha\beta}\nonumber\\
g_{00}^{(1)}&=&0\;,\;g_{0\alpha}^{(1)}=0\;,\;g_{\alpha\beta}^{(1)}=-a^2(t)\left(F\delta_{\alpha\beta}+\frac{9}{10}a(t)t_0^2F_{,\alpha,\beta}\right)\label{e2.1a}\\
g_{00}^{(2)}&=&0\;,\;g_{0\alpha}^{(2)}=0\nonumber\\
g_{\alpha\beta}^{(2)}&=&-a^2(t)\left\{
\frac{9}{40
}a(t)t_0^2\left(-6F_{,\alpha}F_{,\beta}-4FF_{,\alpha,\beta}+F^{,\gamma}F_{,\gamma}\delta_{\alpha\beta}\right)\right.\nonumber\\
\left.\right.&+&\left.\frac{81}{2800}a^2(t)t_0^4\left[19F^{,\gamma}_{,\alpha}F_{,\gamma,\beta}-12F^{,\gamma}F_{,\gamma}F_{,\alpha,\beta}+3\left(\left(F^{,\gamma}_{,\gamma}\right)^2-F^{,\gamma}_{,\nu}F^{,\nu}_{,\gamma}\right)\delta_{\alpha\beta}\right]\right\}
\nonumber
\end{eqnarray}
Here $t_0$ stands for the present time, while $a(t)=(t/t_0)^{2/3}$ denotes the
standard Friedmann-Robertson-Walker scale factor in a matter dominated
homogeneous, isotropic, flat universe. Indices are raised and lowered by the
Eucledian metric $\delta_{\alpha\beta}$.
Our consideration is restricted to scalar perturbations and the expanding
mode.
It is characterized by the scalar function $F(\{x^\alpha\})$ which depends
only on the spatial coordinates.
\begin{eqnarray}
\Gamma_{jk}^i=\Gamma_{jk}^{i(0)}+\Gamma_{jk}^{i(1)}+\Gamma_{jk}^{i(2)}\label{e2.2}
\end{eqnarray}
\begin{eqnarray}
R_{ij}=R_{ij}^{(0)}+R_{ij}^{(1)}+R_{ij}^{(2)}\label{e2.3}
\end{eqnarray}
\begin{eqnarray}
C_{ijkl}=C_{ijkl}^{(1)}+C_{ijkl}^{(2)}\label{e2.4}
\end{eqnarray}
\begin{eqnarray}
u^{i}=u^{i(0)}+u^{i(1)}+u^{i(2)}\label{e2.5}
\end{eqnarray}
\begin{eqnarray}
w^{i}=w^{i(0)}+w^{i(1)}+w^{i(2)}\label{e2.6}
\end{eqnarray}
\begin{eqnarray}
L^{i}_\alpha=L^{i(0)}_\alpha+L^{i(1)}_\alpha+L^{i(2)}_\alpha\label{e2.7}
\end{eqnarray}
\begin{eqnarray}
\Theta=\Theta^{(0)}+\Theta^{(1)}+\Theta^{(2)}\label{e2.8}
\end{eqnarray}
\begin{eqnarray}
w_{\alpha \beta}=w_{\alpha \beta}^{(1)}+w_{\alpha \beta}^{(2)}\label{e2.9}
\end{eqnarray}
\begin{eqnarray}
A=A^{(0)}+A^{(1)}+A^{(2)}\label{e2.10}
\end{eqnarray}
The trajectory of the light ray will be given parametrically as
$\left(t(\lambda), r(\lambda),\vartheta(\lambda),\phi(\lambda)\right)$, which will also be expanded as
\begin{eqnarray}
t(\lambda)=t^{(0)}(\lambda)+t^{(1)}(\lambda)+t^{(2)}(\lambda)\;.\label{e2.10a}
\end{eqnarray}
\begin{eqnarray}
r(\lambda)=r^{(0)}(\lambda)+r^{(1)}(\lambda)+r^{(2)}(\lambda)\;.\label{e2.11}
\end{eqnarray}
\begin{eqnarray}
\vartheta(\lambda)=\vartheta^{(0)}+\vartheta^{(1)}(\lambda)+\vartheta^{(2)}(\lambda)\;.\label{e2.12}
\end{eqnarray}
\begin{eqnarray}
\phi(\lambda)=\phi^{(0)}+\phi^{(1)}(\lambda)+\phi^{(2)}(\lambda)\;.\label{e2.13}
\end{eqnarray}
Note that we shall use throughout a comoving synchronous gauge, since we
consider a dust-filled universe. Hence, $g_{00}=1$ and $g_{0\alpha}=0$ to all
orders (using $c=1$).

As we consider the perturbation of a light ray, all quantities depend on the
parameter $\lambda$ (we shall use mainly $t^{(0)}$ instead). In case of the
field variables this means explicitly that e.g.
\begin{eqnarray}
g_{ij}=g_{ij}(x^j(\lambda))\;,\label{e2.14}
\end{eqnarray}
i.e., we take the arguments along the light ray. This dependence should also
be taken account when performing the perturbation expansion.
\section{Zeroth order}
In zeroth order we write for the line element
\begin{eqnarray}
ds^2=dt^2-a^2(t)\left(dr^2+r^2\left(d\vartheta^2+\sin^2 \vartheta d\phi^2\right)\right)\label{e3.1}
\end{eqnarray}
thus $g_{ij}^{(0)}$ is diagonal, and
\begin{eqnarray}
g_{tt}^{(0)}=1\label{e3.2}\\
g_{rr}^{(0)}=-a^2(t)\label{e3.3}\\
g_{\vartheta\vartheta}^{(0)}=-a^2(t)r^2\label{e3.4}\\
g_{\phi\phi}^{(0)}=-a^2(t)r^2\sin^2 \vartheta\label{e3.5}
\end{eqnarray}
In a matter dominated universe with zero pressure we have
\begin{eqnarray}
a(t)=a_0\left(\frac{t-\tau}{t_0-\tau}\right)^{\frac{2}{3}}\label{e3.5a}
\end{eqnarray}
where $a_0$ stands for the scale factor at the present age $t_0$ and $\tau$ is a parameter depending on the previous 
history of the universe. In what follows, we set $\tau=0$, which means that $t$ is not measured from the Big Bang,
and thus $t_0$ is not the actual age of the universe. Certainly, one may change in the final expressions $t$ to 
$t-\tau$ again. Also, we take $a_0=1$.
The nonzero components of the Christoffel symbol are
\begin{eqnarray}
\Gamma_{rr}^{t(0)}=a\dot a\label{e3.6}\\
\Gamma_{\vartheta\vartheta}^{t(0)}=r^2a\dot a\label{e3.7}\\
\Gamma_{\phi\phi}^{t(0)}=r^2\sin^2 \vartheta a\dot a\label{e3.8}\\
\Gamma_{tr}^{r(0)}=\Gamma_{rt}^{r(0)}=\Gamma_{t\vartheta}^{\vartheta(0)}=\Gamma_{\vartheta
  t}^{\vartheta(0)}=\Gamma_{t\phi}^{\phi(0)}=\Gamma_{\phi t}^{\phi(0)}=\frac{\dot a}{a}\label{e3.9}\\
\Gamma_{\vartheta\vartheta}^{r(0)}=-r\label{e3.10}\\
\Gamma_{\phi\phi}^{r(0)}=-r\sin^2 \vartheta\label{e3.11}\\
\Gamma_{\phi\phi}^{\vartheta(0)}=-\sin \vartheta \cos \vartheta\label{e3.12}\\
\Gamma_{\vartheta r}^{\vartheta(0)}=\Gamma_{r \vartheta}^{\vartheta(0)}=\Gamma_{\phi
  r}^{\phi(0)}=\Gamma_{r \phi}^{\phi(0)}=\frac{1}{r}\label{e3.13}\\
\Gamma_{\vartheta\phi}^{\phi(0)}=\Gamma_{\phi\vartheta}^{\phi(0)}=\cot \vartheta\label{e3.14}
\end{eqnarray}
The Ricci tensor is diagonal with the elements
\begin{eqnarray}
R_{tt}^{(0)}=-3\frac{\ddot a}{a}\label{e3.15}\\
R_{rr}^{(0)}=a\ddot a+2\dot a^2\label{e3.16}\\
R_{\vartheta\vartheta}^{(0)}=r^2\left(a\ddot a+2\dot a^2\right)\label{e3.17}\\
R_{\phi\phi}^{(0)}=r^2\sin^2 \vartheta\left(a\ddot a+2\dot a^2\right)\label{e3.18}
\end{eqnarray}
The Weyl tensor vanishes.
The results for a light beam emitted from a point source at time $t_1$  
and arriving at the origin at $t_0$ are the following:
\begin{eqnarray}
u^{t(0)}=\frac{1}{a}\label{e3.19}\\
u^{r(0)}=-\frac{1}{a^2}\label{e3.20}\\
u^{\vartheta(0)}=u^{\phi(0)}=0\label{e3.21}
\end{eqnarray}
For the trajectory we get
\begin{eqnarray}
\dot r^{(0)}=-\frac{1}{a}\label{e3.22}
\end{eqnarray}
or
\begin{eqnarray}
r^{(0)}(t)=-\int_{t_0}^{t} \frac{dt}{a}=3\left(\frac{t_0}{a(t_0)}-\frac{t}{a(t)}\right)\label{e3.23}
\end{eqnarray}
while $\vartheta^{(0)}$ and $\phi^{(0)}$ are constant.
We also introduce the radial coordinate distance from the source:
\begin{eqnarray}
d(t)=r^{(0)}(t_1)-r^{(0)}(t)=3\left(\frac{t}{a(t)}-\frac{t_1}{a(t_1)}\right)\;.\label{e3.23a}
\end{eqnarray}
For the expansion rate we have
\begin{eqnarray}
\Theta^{(0)}=\frac{\dot a}{a^2}+\frac{1}{a^2 d}=\frac{\dot{(a d)}}{a^2 d}\label{e3.24}
\end{eqnarray}
and thus the cross section area of the beam is
\begin{eqnarray}
A^{(0)}=\Omega a^2d^2\label{e3.25}
\end{eqnarray}
The intensity of the light beam is then 
\begin{eqnarray}
I^{(0)}=L\frac{\Omega}{4\pi}\frac{u^{0(0)2}(t_0)}{u^{0(0)2}(t_1)A^{(0)}(t_0)}
=L\frac{a^2(t_1)}{4\pi a^4(t_0)d^2(t_0)}\label{e3.25a}
\end{eqnarray}
where $L$ stands for the absolute luminosity (radiated energy pro time unit). 
It is customary to express it in terms of the red shift 
\begin{eqnarray}
z=\frac{\rm measured\;wavelength - emitted\;wavelength}{\rm emitted\;wavelength}=
\frac{u^{0(0)}(t_1)}{u^{0(0)}(t_0)}-1
=\frac{a(t_0)}{a(t_1)}-1\label{e3.25b}
\end{eqnarray}
Since the scale factor $a$ is proportional to $t^{2/3}$, we have
\begin{eqnarray}
I^{(0)}=\frac{L}{4\pi}\frac{H_0^2}{4}\frac{1}{(1+z)\left(\sqrt{1+z}-1\right)^2}
=\frac{L}{4\pi}H_0^2\left(\frac{1}{z^2}-\frac{1}{2z}+\frac{7}{16}-\frac{13}{32}z+...\right)\label{e3.25c}
\end{eqnarray}
or 
\begin{eqnarray}
-\log I^{(0)}=-\log{\left(\frac{LH_0^2}{4\pi}\right)} +2\log{z} 
+\frac{1}{2}z-\frac{7}{16}z^2+\frac{11}{48}z^3+...\label{e3.25d}
\end{eqnarray}
The Hubble parameter $H_0$ is given by
\begin{eqnarray}
H_0=\left.\frac{\dot a}{a}\right|_{t=t_0}=\frac{2}{3 t_0}\label{e3.25e}
\end{eqnarray}
Note again that $t_0$ is not the actual age of the universe.

We shall also need the zeroth order expression of $w^i$ and $L^i_\alpha$. They are given by
\begin{eqnarray}
w^{t(0)}=\frac{a}{2}\label{e3.26}\\
w^{r(0)}=\frac{1}{2}\label{e3.27}\\
w^{\vartheta(0)}=w^{\phi(0)}=0\label{e3.28}
\end{eqnarray}
and
\begin{eqnarray}
L^{t(0)}_{1,2}=L^{r(0)}_{1,2}=0\label{e3.29}\\
L^{\vartheta(0)}_1=\frac{1}{ar^{(0)}}\label{e3.30}\\
L^{\phi(0)}_1=0\label{e3.31}\\
L^{\vartheta(0)}_2=0\label{e3.32}\\
L^{\phi(0)}_2=\frac{1}{ar^{(0)}\sin \vartheta^{(0)}}\label{e3.33}
\end{eqnarray}
\section{First order}
As $w_{\alpha \beta}$ vanishes in zeroth order, Eq.(\ref{e1.17}) implies that
for a second order calculation of $\Theta$ one needs the first order
corrections of  $w_{\alpha \beta}$ and $C_{ijkl}$.
In contrast, corrections of $R_{jk}$ and $u^j$ are needed up to second order.
As the corrections to the metric are given, we consider it and all of its
derivatives as known quantities and seek for corrections of the path,
expansion and shear of the light beam.  
In this section we consider the first order corrections. As the Sachs optical
equations refer to a certain light trajectory, the deformation of that
trajectory should also be taken into account. Therefore, we start with the
equations (\ref{e1.1})-(\ref{e1.3}). 
For the first order correction of $u^j$
we have
\begin{eqnarray}
2g^{(0)}_{ij}u^{i(1)}u_j^{j(0)}+g^{(1)}_{ij}u^{i(0)}u^{j(0)}+\frac{\partial g_{jk}^{(0)}}{\partial x^l}x^{l(1)}u^{j(0)}u^{k(0)}=0\;.
\label{e4.1}
\end{eqnarray}
Inserting the zeroth order expressions we have 
\begin{eqnarray}
\frac{2}{a}\frac{dt^{(1)}}{d\lambda}+2\frac{dr^{(1)}}{d\lambda}+\frac{g^{(1)}_{rr}}{a^4}-\frac{2\dot a}{a^3}t^{(1)}=0\;.
\label{e4.1a}
\end{eqnarray} 
Similarly, the geodesic equation implies
\begin{eqnarray}
\frac{du^{i(1)}}{d\lambda}+2\Gamma^{i(0)}_{jk}u^{j(1)}u^{k(0)}+\Gamma^{i(1)}_{jk}u^{j(0)}u^{k(0)}
+\frac{\partial \Gamma^{i(0)}_{jk}}{\partial x^l}x^{l(1)}u^{j(0)}u^{k(0)}=0\;.
\label{e4.2}
\end{eqnarray}
Inserting the zeroth order expressions this leads to 
\begin{eqnarray}
\frac{d^2t^{(1)}}{d\lambda^2}-2\frac{\dot a}{a}\frac{dr^{(1)}}{d\lambda}
+\left(\frac{\ddot a}{a^3}+\frac{\dot
  a^2}{a^4}\right)t^{(1)}=-\Gamma^{t(1)}_{jk}u^{j(0)}u^{k(0)}\label{e4.2a}\\
\frac{d^2r^{(1)}}{d\lambda^2}-2\frac{\dot a}{a^3}\frac{dt^{(1)}}{d\lambda}+2\frac{\dot a}{a^2}\frac{dr^{(1)}}{d\lambda}
-2\left(\frac{\ddot a}{a^4}-\frac{\dot
  a^2}{a^5}\right)t^{(1)}=-\Gamma^{r(1)}_{jk}u^{j(0)}u^{k(0)}\label{e4.2b}\\
\frac{d^2\vartheta^{(1)}}{d\lambda^2}+2\frac{\dot{(ar^{(0)})}}{a^2r^{(0)}}\frac{d\vartheta^{(1)}}{d\lambda}
=-\Gamma^{\vartheta(1)}_{jk}u^{j(0)}u^{k(0)}\label{e4.2c}\\
\frac{d^2\phi^{(1)}}{d\lambda^2}+2\frac{\dot{(ar^{(0)})}}{a^2r^{(0)}}\frac{d\phi^{(1)}}{d\lambda}
=-\Gamma^{\phi(1)}_{jk}u^{j(0)}u^{k(0)}\label{e4.2d}
\end{eqnarray} 
As the right hand sides of these equations are in principle known functions of
$\lambda$ (or, equivalently, of $\zeta\equiv t^{(0)}$, cf. Eq.(\ref{e3.19})) Eqs.(\ref{e4.2c}), (\ref{e4.2d})
can be directly integrated to get
\begin{eqnarray}
u^{\vartheta(1)}(\zeta)=-\frac{1}{a^2r^{(0)2}}\int_{t_0}^{\zeta}d\zeta'\;a^3r^{(0)2}\;\Gamma^{\vartheta(1)}_{jk}u^{j(0)}u^{k(0)}
\label{e4.3}
\end{eqnarray}  
Here the constant of integration (or, equivalently, the lower limit of the integral) is determined by the criterion that 
$u^{\vartheta(1)}(\zeta)$ must not diverge at $\zeta=t_0$. A divergence corresponds to a zeroth order change in the 
direction, describing a light ray not hitting the observation point (i.e., the origin). As Eq.(\ref{e4.3}) shows, such 
divergence does not occur, and the direction of the light ray is not even changed at $\zeta=t_0$. It does change 
(due to the perturbation), however, at the source, i.e., at  $\zeta=t_1$. 

A further integration yields the first correction to the angle $\vartheta$,  namely,
\newpage
\begin{eqnarray} 
\vartheta^{(1)}(\zeta)&=&-\frac{1}{r^{(0)}(\zeta)}\int_{t_0}^{\zeta}d\zeta'\;a^3r^{(0)2}\;\Gamma^{\vartheta(1)}_{jk}u^{j(0)}u^{k(0)}\nonumber\\
&+&\frac{1}{r^{(0)}(t_1)}\int_{t_0}^{t_1}d\zeta'\;a^3r^{(0)2}\;\Gamma^{\vartheta(1)}_{jk}u^{j(0)}u^{k(0)}\label{e4.4}\\
&+&\int_{t_1}^{\zeta}d\zeta'\;a^3r^{(0)}\;\Gamma^{\vartheta(1)}_{jk}u^{j(0)}u^{k(0)}
\nonumber
\end{eqnarray}  
The integration constant has been 
determined by the condition that $\vartheta^{(1)}(t_1)=0$, as the initial position is fixed.
Similarly we have for the other angle
\begin{eqnarray}
u^{\phi(1)}(\zeta)=-\frac{1}{a^2r^{(0)2}}\int_{t_0}^{\zeta}d\zeta'\;a^3r^{(0)2}\;\Gamma^{\phi(1)}_{jk}u^{j(0)}u^{k(0)}
\label{e4.5}
\end{eqnarray}
\begin{eqnarray} 
\phi^{(1)}(\zeta)&=&-\frac{1}{r^{(0)}(\zeta)}\int_{t_0}^{\zeta}d\zeta'\;a^3r^{(0)2}\;\Gamma^{\phi(1)}_{jk}u^{j(0)}u^{k(0)}\nonumber\\
&+&\frac{1}{r^{(0)}(t_1)}\int_{t_0}^{t_1}d\zeta'\;a^3r^{(0)2}\;\Gamma^{\phi(1)}_{jk}u^{j(0)}u^{k(0)}\label{e4.6}\\
&+&\int_{t_1}^{\zeta}d\zeta'\;a^3r^{(0)}\;\Gamma^{\phi(1)}_{jk}u^{j(0)}u^{k(0)}
\nonumber
\end{eqnarray}   
All quantities above are considered as functions of $\zeta\equiv t^{(0)}$. 
  
Eqs.(\ref{e4.1a}), (\ref{e4.2a}), (\ref{e4.2b}) are not independent, hence, it is advisable to use Eqs.(\ref{e4.1a}) 
and (\ref{e4.2a}). Then, expressing $\frac{dr^{(1)}}{d\lambda}$ from Eq.(\ref{e4.1a}) and putting it into Eq.(\ref{e4.2a})
we have
\begin{eqnarray}
\frac{1}{a^2}\ddot t^{(1)} +\frac{\dot a}{a^3}\dot t^{(1)}
+\left(\frac{\ddot a}{a^3}-\frac{\dot a^2}{a^4}\right)t^{(1)}
=-\frac{\dot a}{a^5}g^{(1)}_{rr}-\Gamma^{t(1)}_{jk}u^{j(0)}u^{k(0)}\label{e4.7}
\end{eqnarray}
Again, a dot means derivative with respect to $\zeta\equiv t^{(0)}$. 
The solution of Eq.(\ref{e4.7}) can be written as
\begin{eqnarray}
t^{(1)}=\frac{3}{5}\zeta\int_{t_0}^{\zeta}d\xi\;f(\xi)-\frac{3}{5}\zeta^{-\frac{2}{3}}
\int_{t_0}^{\zeta}d\xi\;\xi^{\frac{5}{3}}f(\xi)+\alpha\zeta+\beta\zeta^{-\frac{2}{3}}
\label{e4.8}
\end{eqnarray}
where
\begin{eqnarray}
f(\zeta)=-\frac{\dot
  a}{a^3}g^{(1)}_{rr}-a^2\Gamma^{t(1)}_{jk}u^{j(0)}u^{k(0)}=-\frac{3}{10}t_0^\frac{4}{3}\zeta^{-\frac{1}{3}}\frac{\partial^2
  F}{\partial r^2}
\label{e4.9}
\end{eqnarray}
while $\alpha$ and $\beta$ are integration constants. Note that in
Eq.(\ref{e4.9}) the arguments of $\partial^2
  F/\partial r^2$ are to be taken along the light ray, hence, it also
  depends on $\zeta$.
We also made use of the fact that $a\propto \zeta^{2/3}$.
Now we have
\begin{eqnarray}
u^{t(1)}=\frac{dt^{(1)}}{d\lambda}=\frac{1}{a}\frac{dt^{(1)}}{d\zeta}\nonumber\\
=\frac{1}{a}\left(\frac{3}{5}\int_{t_0}^{\zeta}d\xi\;f(\xi)+\frac{2}{5}\zeta^{-\frac{5}{3}}
\int_{t_0}^{\zeta}d\xi\;\xi^{\frac{5}{3}}f(\xi)+\alpha-\frac{2}{3}\beta\zeta^{-\frac{5}{3}}\right)\;,
\label{e4.10}
\end{eqnarray}
\begin{eqnarray}
u^{r(1)}=-\frac{1}{a^2}\left(\frac{1}{5}\int_{t_0}^{\zeta}d\xi\;f(\xi)
+\frac{4}{5}\zeta^{-\frac{5}{3}}\int_{t_0}^{\zeta}d\xi\;\xi^{\frac{5}{3}}f(\xi)
+\frac{g^{(1)}_{rr}}{2a^2}+\frac{1}{3}\alpha-\frac{4}{3}\beta\zeta^{-\frac{5}{3}}\right)
\label{e4.11}
\end{eqnarray}
and
\begin{eqnarray}
r^{(1)}=-\frac{3}{5}\frac{\zeta}{a}\int_{t_0}^{\zeta}d\xi\;f(\xi)
+\frac{3}{5}\frac{\zeta^{-\frac{2}{3}}}{a}\int_{t_0}^{\zeta}d\xi\;\xi^{\frac{5}{3}}f(\xi)
-\int_{t_0}^{\zeta}d\xi\;\frac{g^{(1)}_{rr}}{2a^3}-\frac{\alpha}{a}\zeta-\frac{\beta}{a}\zeta^{-\frac{2}{3}}+\gamma
\label{e4.12}
\end{eqnarray}
where $\gamma$ is a further integration constant. The three integration constants $\alpha$, $\beta$, $\gamma$ are
determined by the conditions
\begin{eqnarray}
t^{(1)}(t_0)=0\label{e4.12a}\\
r^{(1)}(t_0)=0\label{e4.12b}\\
t^{(1)}(t_1)=0\label{e4.12c}
\end{eqnarray}
We get
\begin{eqnarray}
\alpha&=&\frac{3}{5}\frac{t_1^{\frac{5}{3}}}{t_0^{\frac{5}{3}}-t_1^{\frac{5}{3}}}
\int_{t_0}^{t_1}d\xi\;\left(1-\left(\frac{\xi}{t_1}\right)^{\frac{5}{3}}\right)f(\xi)
\label{e4.12d}\\
\beta&=&-\frac{3}{5}\frac{t_1^{\frac{5}{3}}t_0^{\frac{5}{3}}}{t_0^{\frac{5}{3}}-t_1^{\frac{5}{3}}}
\int_{t_0}^{t_1}d\xi\;\left(1-\left(\frac{\xi}{t_1}\right)^{\frac{5}{3}}\right)f(\xi)
\label{e4.12e}\\
\gamma&=&0\label{e4.12f}
\end{eqnarray}
The first order correction of the expansion rate $\Theta$ satisfies the equation (cf. Eq.(\ref{e1.17}))
\begin{eqnarray}
\frac{d\Theta^{(1)}}{d\lambda}+2\Theta^{(0)}\Theta^{(1)}=-\frac{1}{2}R_{jk}^{(1)}u^{j(0)}u^{k(0)}
-R_{jk}^{(0)}u^{j(1)}u^{k(0)}-\frac{1}{2}\frac{\partial R_{jk}^{(0)}}{\partial x^l}x^{l(1)}u^{j(0)}u^{k(0)}
\label{e4.13}
\end{eqnarray}
Inserting the zeroth order terms on the left hand side, we have
\begin{eqnarray}
\frac{1}{a}\frac{d\Theta^{(1)}}{d\zeta}+2\frac{\dot{(a d)}}{a^2 d}\Theta^{(1)}=\chi^{(1)}(\zeta)
\label{e4.14}
\end{eqnarray}
where $\chi^{(1)}(\zeta)$ stands for the right hand side of
Eq.(\ref{e4.13}), that is a known function by now. Explicitly, it is
\begin{eqnarray}
\chi^{(1)}(\zeta)
&=&\frac{3}{10}\left(\frac{\zeta}{t_0}\right)^{-\frac{8}{3}}
\triangle F\label{e4.14a}\\
&-&\frac{2}{5}\frac{1}{t_0}\left(\frac{\zeta}{t_0}\right)^{-3}\frac{\partial F}{\partial r}-\frac{4}{5}\frac{1}{t_0^2}\left(\frac{\zeta}{t_0}\right)^{-\frac{10}{3}}F+\frac{4}{3}\frac{1}{t_0^3}\left(\frac{\zeta}{t_0}\right)^{-5}\int_{t_0}^{\zeta}d\xi\;\left(\frac{\xi}{t_0}\right)^{\frac{2}{3}}F\nonumber\\&+&\frac{2}{3}\frac{1}{t_0^2}\left(\frac{\zeta}{t_0}\right)^{-5}\frac{F(t_1)-F(t_0)}{\left(\frac{t_0}{t_1}\right)^{\frac{5}{3}}-1}-\frac{4}{3}\frac{1}{t_0^3}\left(\frac{\zeta}{t_0}\right)^{-5}\frac{\left(\frac{t_0}{t_1}\right)^{\frac{5}{3}}}{\left(\frac{t_0}{t_1}\right)^{\frac{5}{3}}-1}\int_{t_0}^{t_1}d\xi\;\left(\frac{\xi}{t_0}\right)^{\frac{2}{3}}F
\nonumber
\end{eqnarray}
The solution of Eq.(\ref{e4.14}) can be readily written as
\begin{eqnarray}
\Theta^{(1)}=\frac{1}{a^2 d^{2}}\int_{t_1}^{\zeta}d\xi\;a^3 d^{2}\chi^{(1)}(\xi)
\label{e4.15}
\end{eqnarray}
Note that the integration constant (i.e., the lower integration border) is
fixed by the condition that no singularity can appear at $\zeta=t_1$ since
that would amount to a change of the light source position.
The first order correction of the beam area satisfies the equation (cf. Eq.(\ref{e1.20}))
\begin{eqnarray}
\frac{dA^{(1)}}{d\lambda}=2\Theta^{(0)}A^{(1)}+2\Theta^{(1)}A^{(0)}
\label{e4.16}
\end{eqnarray}
or, inserting the explicit zeroth order expressions 
\begin{eqnarray}
\frac{1}{a}\frac{dA^{(1)}}{d\zeta}-2\frac{\dot{(ad)}}{a^2 d}A^{(1)}=2\Theta^{(1)}\Omega a^2d^{2}\;.
\label{e4.17}
\end{eqnarray}
Its solution is
\begin{eqnarray}
A^{(1)}=2\Omega a^2 d^{2}
\int_{t_1}^{\zeta}d\xi\;a \Theta^{(1)}=2\Omega a^2 d^{2}\left(
-\frac{1}{d}\int_{t_1}^{\zeta}d\xi\;a^3 d^{2}\chi^{(1)}(\xi)\right.\nonumber\\\left.
+\int_{t_1}^{\zeta}d\xi\;a^3 d\;\chi^{(1)}(\xi)
\right)\;.
\label{e4.18}
\end{eqnarray}
The integration constant is determined by the condition that the solid angle at small distances is given,
i.e., $A^{(1)}/(a^2 d^{2})$ vanishes when $\zeta\rightarrow t_1$.
Explicitly we have
\begin{eqnarray}
A^{(1)}&=&A^{(0)}\left\{\frac{1}{5}\frac{\left(t_1^{-\frac{1}{3}}-t_0^{-\frac{1}{3}}\right)\left(t_1^{-\frac{2}{3}}-t_0^{-\frac{2}{3}}\right)\left(2t_1^{-\frac{2}{3}}+t_1^{-\frac{1}{3}}t_0^{-\frac{1}{3}}+2t_0^{-\frac{2}{3}}\right)}{t_1^{-\frac{5}{3}}-t_0^{-\frac{5}{3}}}\left(F(t_1)-F(t_0)\right)\right.\nonumber\\&+&\left.2\frac{t_1^{-\frac{5}{3}}t_0^{-\frac{5}{3}}}{t_1^{-\frac{5}{3}}-t_0^{-\frac{5}{3}}}\int_{t_1}^{t_0}d\xi\;\xi^{\frac{2}{3}}F(\xi)-\frac{2}{5}\frac{1}{t_0^{\frac{1}{3}}-t_1^{\frac{1}{3}}}\int_{t_1}^{t_0}d\xi\;\xi^{-\frac{2}{3}}F(\xi)\right.\nonumber\\&+&\left.\frac{9}{5}\frac{t_0^{\frac{4}{3}}}{t_0^{\frac{1}{3}}-t_1^{\frac{1}{3}}}\int_{t_1}^{t_0}d\xi\;\triangle
F(\xi) \left(-1+(t_0^{\frac{1}{3}}+t_1^{\frac{1}{3}})\xi^{-\frac{1}{3}}-t_0^{\frac{1}{3}}t_1^{\frac{1}{3}}\xi^{-\frac{2}{3}}\right)\right\}
\;.
\label{e4.18_expl}
\end{eqnarray}
Let us sketch how to get the corrections to the Hubble diagram. We have now expressions for $A$ and $u^t$ parametrized by 
$t_1$. (Note that $t_0$ is fixed.)  The intensity and redshift are expressed as
\begin{eqnarray}
I=L\frac{\Omega}{4\pi}\frac{\left(u^{t}(t_0)\right)^2}{\left(u^{t}(t_1)\right)^2A(t_0)}
\label{e4.18a}
\end{eqnarray}
and
\begin{eqnarray}
z=\frac{u^t(t_1)}{u^t(t_0)}-1\;,
\label{e4.18b}
\end{eqnarray}
respectively. Hence, the Hubble diagram $I(z)$ is given 
in a parametric form.

In the second order calculation we shall also need the first order correction to $w_{\alpha\beta}$.
From Eq.(\ref{e1.18}) we get 
\begin{eqnarray}
\frac{dw_{\alpha\beta}^{(1)}}{d\lambda}+2\Theta^{(0)} w_{\alpha\beta}^{(1)} =C_{ ijkl}^{(1)}L^{i(0)}_\alpha
u^{j(0)}u^{k(0)}L^{l(0)}_\beta
\label{e4.19}
\end{eqnarray}
i.e.
\begin{eqnarray}
\frac{1}{a}\frac{dw_{\alpha\beta}^{(1)}}{d\zeta}+2\frac{\dot{(ad)}}{a^2 d} w_{\alpha\beta}^{(1)} =C_{ ijkl}^{(1)}L^{i(0)}_\alpha u^{j(0)}u^{k(0)}L^{l(0)}_\beta\;.
\label{e4.20}
\end{eqnarray}
The solution sounds
\begin{eqnarray}
w_{\alpha\beta}^{(1)}=\frac{1}{a^2 d^{2}}\int_{t_1}^{\zeta}d\xi\;a^3 d^{2}C_{ ijkl}^{(1)}L^{i(0)}_\alpha
u^{j(0)}u^{k(0)}L^{l(0)}_\beta\;.
\label{e4.21}
\end{eqnarray}
Again, the integration constant is fixed by the condition that no singularity
may appear at $\zeta=t_1$.

For completeness, below we present the explicit expressions for all the relevant
first order corrections.
\begin{eqnarray}
t^{(1)}(\zeta)&=&-\frac{3}{10}\zeta \;F(\zeta)\\
&+&\frac{3}{10}t_0\left(\frac{t_1}{t_0}\right)^{\frac{5}{3}}\left(\frac{\zeta}{t_0}\right)^{-\frac{2}{3}}\frac{t_0^{\frac{5}{3}}-\zeta^{\frac{5}{3}}}{t_0^{\frac{5}{3}}-t_1^{\frac{5}{3}}}F(t_1)+\frac{3}{10}t_0\left(\frac{\zeta}{t_0}\right)^{-\frac{2}{3}}\frac{\zeta^{\frac{5}{3}}-t_1^{\frac{5}{3}}}{t_0^{\frac{5}{3}}-t_1^{\frac{5}{3}}}F(t_0)
\nonumber\\
&+&\frac{3}{5}\left(\frac{\zeta}{t_0}\right)^{-\frac{2}{3}}\int_{t_0}^{\zeta}d\xi\;\left(\frac{\xi}{t_0}\right)^{\frac{2}{3}}F(\xi)-\frac{3}{5}\left(\frac{\zeta}{t_0}\right)^{-\frac{2}{3}}\frac{t_0^{\frac{5}{3}}-\zeta^{\frac{5}{3}}}{t_0^{\frac{5}{3}}-t_1^{\frac{5}{3}}}\int_{t_0}^{t_1}d\xi\;\left(\frac{\xi}{t_0}\right)^{\frac{2}{3}}F(\xi)
\nonumber\\
r^{(1)}(\zeta)&=&-\frac{3}{5}\int_\zeta^{t_0}
d\xi\;\left(\frac{\xi}{t_0}\right)^{-\frac{2}{3}}F(\xi)\label{e4.22}\\&+&\frac{9}{20}t_0^2
\left(\frac{\partial F}{\partial
  r}(t_0)-\left(\frac{\zeta}{t_0}\right)^{\frac{2}{3}}\frac{\partial
  F}{\partial r}(\zeta)\right)
+\frac{3}{10}t_0\left(F(t_0)-\left(\frac{\zeta}{t_0}\right)^{\frac{1}{3}}F(\zeta)\right)\nonumber\\&+&t_0\left(\frac{\zeta}{t_0}\right)^{\frac{1}{3}}\left(1-\left(\frac{\zeta}{t_0}\right)^{-\frac{5}{3}}\right)\left[\frac{3}{10}\frac{t_0^{\frac{5}{3}}F(t_0)-\zeta^{\frac{5}{3}}F(\zeta)}{t_0^{\frac{5}{3}}-\zeta^{\frac{5}{3}}}-\frac{3}{10}\frac{t_0^{\frac{5}{3}}F(t_0)-t_1^{\frac{5}{3}}F(t_1)}{t_0^{\frac{5}{3}}-t_1^{\frac{5}{3}}}\right.\nonumber\\&-&\left.\frac{3}{5}\frac{1}{t_0}\frac{1}{1-\left(\frac{\zeta}{t_0}\right)^{\frac{5}{3}}}\int_\zeta^{t_0}d\xi\;\left(\frac{\xi}{t_0}\right)^{\frac{2}{3}}F(\xi)+\frac{3}{5}\frac{1}{t_0}\frac{1}{1-\left(\frac{t_1}{t_0}\right)^{\frac{5}{3}}}\int_{t_1}^{t_0}d\xi\;\left(\frac{\xi}{t_0}\right)^{\frac{2}{3}}F(\xi)\right]
\nonumber\\
\vartheta^{(1)}(\zeta)&=&-\frac{1}{r^{(0)}(\zeta)}\int_{t_0}^{\zeta}d\xi\;\left[\frac{9}{20}\frac{t_0^2}{r^{(0)2}}\left(\frac{\partial
  F}{\partial
  \vartheta}-r^{(0)}\frac{\partial^2
  F}{\partial r\partial
  \vartheta}+\frac{r^{(0)2}}{2}\frac{\partial^3
  F}{\partial^2 r\partial
  \vartheta}\right)\right.\nonumber\\
&+&\left.\frac{3}{5}\frac{t_0}{r^{(0)}}\left(\frac{\xi}{t_0}\right)^{-\frac{1}{3}}\left(\frac{\partial
  F}{\partial
  \vartheta}-r^{(0)}\frac{\partial^2
  F}{\partial r\partial
  \vartheta}\right)-\frac{1}{2}\left(\frac{\xi}{t_0}\right)^{-\frac{2}{3}}\frac{\partial
  F}{\partial
  \vartheta}\right]\nonumber\\
&+&\frac{1}{r^{(0)}(t_1)}\int_{t_0}^{t_1}d\xi\;\left[\frac{9}{20}\frac{t_0^2}{r^{(0)2}}\left(\frac{\partial
  F}{\partial
  \vartheta}-r^{(0)}\frac{\partial^2
  F}{\partial r\partial
  \vartheta}+\frac{r^{(0)2}}{2}\frac{\partial^3
  F}{\partial^2 r\partial
  \vartheta}\right)\right.\nonumber\\
&+&\left.\frac{3}{5}\frac{t_0}{r^{(0)}}\left(\frac{\xi}{t_0}\right)^{-\frac{1}{3}}\left(\frac{\partial
  F}{\partial
  \vartheta}-r^{(0)}\frac{\partial^2
  F}{\partial r\partial
  \vartheta}\right)-\frac{1}{2}\left(\frac{\xi}{t_0}\right)^{-\frac{2}{3}}\frac{\partial
  F}{\partial
  \vartheta}\right]\nonumber\\
&+&\int_{t_1}^{\zeta}d\xi\;\left[\frac{9}{20}\frac{t_0^2}{r^{(0)3}}\left(\frac{\partial
  F}{\partial
  \vartheta}-r^{(0)}\frac{\partial^2
  F}{\partial r\partial
  \vartheta}+\frac{r^{(0)2}}{2}\frac{\partial^3
  F}{\partial^2 r\partial
  \vartheta}\right)\right.\nonumber\\
&+&\left.\frac{3}{5}\frac{t_0}{r^{(0)2}}\left(\frac{\xi}{t_0}\right)^{-\frac{1}{3}}\left(\frac{\partial
  F}{\partial
  \vartheta}-r^{(0)}\frac{\partial^2
  F}{\partial r\partial
  \vartheta}\right)-\frac{1}{2}\frac{1}{r^{(0)}}\left(\frac{\xi}{t_0}\right)^{-\frac{2}{3}}\frac{\partial
  F}{\partial
  \vartheta}\right]\\
\phi^{(1)}(\zeta)&=&-\frac{1}{r^{(0)}(\zeta)\sin^2\vartheta}\int_{t_0}^{\zeta}d\xi\;\left[\frac{9}{20}\frac{t_0^2}{r^{(0)2}}\left(\frac{\partial
  F}{\partial
  \phi}-r^{(0)}\frac{\partial^2
  F}{\partial r\partial
  \phi}+\frac{r^{(0)2}}{2}\frac{\partial^3
  F}{\partial^2 r\partial
  \phi}\right)\right.\nonumber\\
&+&\left.\frac{3}{5}\frac{t_0}{r^{(0)}}\left(\frac{\xi}{t_0}\right)^{-\frac{1}{3}}\left(\frac{\partial
  F}{\partial
  \phi}-r^{(0)}\frac{\partial^2
  F}{\partial r\partial
  \phi}\right)-\frac{1}{2}\left(\frac{\xi}{t_0}\right)^{-\frac{2}{3}}\frac{\partial
  F}{\partial
  \phi}\right]\nonumber\\
&+&\frac{1}{r^{(0)}(t_1)\sin^2\vartheta}\int_{t_0}^{t_1}d\xi\;\left[\frac{9}{20}\frac{t_0^2}{r^{(0)2}}\left(\frac{\partial
  F}{\partial
  \phi}-r^{(0)}\frac{\partial^2
  F}{\partial r\partial
  \phi}+\frac{r^{(0)2}}{2}\frac{\partial^3
  F}{\partial^2 r\partial
  \phi}\right)\right.\nonumber\\
&+&\left.\frac{3}{5}\frac{t_0}{r^{(0)}}\left(\frac{\xi}{t_0}\right)^{-\frac{1}{3}}\left(\frac{\partial
  F}{\partial
  \phi}-r^{(0)}\frac{\partial^2
  F}{\partial r\partial
  \phi}\right)-\frac{1}{2}\left(\frac{\xi}{t_0}\right)^{-\frac{2}{3}}\frac{\partial
  F}{\partial
  \phi}\right]\nonumber\\
&+&\frac{1}{\sin^2\vartheta}\int_{t_1}^{\zeta}d\xi\;\left[\frac{9}{20}\frac{t_0^2}{r^{(0)3}}\left(\frac{\partial
  F}{\partial
  \phi}-r^{(0)}\frac{\partial^2
  F}{\partial r\partial
  \phi}+\frac{r^{(0)2}}{2}\frac{\partial^3
  F}{\partial^2 r\partial
  \phi}\right)\right.\nonumber\\
&+&\left.\frac{3}{5}\frac{t_0}{r^{(0)2}}\left(\frac{\xi}{t_0}\right)^{-\frac{1}{3}}\left(\frac{\partial
  F}{\partial
  \phi}-r^{(0)}\frac{\partial^2
  F}{\partial r\partial
  \phi}\right)-\frac{1}{2}\frac{1}{r^{(0)}}\left(\frac{\xi}{t_0}\right)^{-\frac{2}{3}}\frac{\partial
  F}{\partial
  \phi}\right]\\
u^{t(1)}(\zeta)&=&\frac{3}{10}t_0\left(\frac{\zeta}{t_0}\right)^{-\frac{1}{3}}\frac{\partial F}{\partial
  r}(\zeta)+\frac{3}{10}\left(\frac{\zeta}{t_0}\right)^{-\frac{2}{3}} \;F(\zeta)\\
&-&\frac{1}{10}\left(\frac{t_1}{t_0}\right)^{\frac{5}{3}}\left(\frac{\zeta}{t_0}\right)^{-\frac{7}{3}}\frac{2t_0^{\frac{5}{3}}+3\zeta^{\frac{5}{3}}}{t_0^{\frac{5}{3}}-t_1^{\frac{5}{3}}}F(t_1)+\frac{1}{10}\left(\frac{\zeta}{t_0}\right)^{-\frac{7}{3}}\frac{2t_1^{\frac{5}{3}}+3\zeta^{\frac{5}{3}}}{t_0^{\frac{5}{3}}-t_1^{\frac{5}{3}}}F(t_0)
\nonumber\\
&-&\frac{2}{5}\frac{1}{t_0}\left(\frac{\zeta}{t_0}\right)^{-\frac{7}{3}}\int_{t_0}^{\zeta}d\xi\;\left(\frac{\xi}{t_0}\right)^{\frac{2}{3}}F(\xi)+\frac{1}{5}\frac{1}{t_0}\left(\frac{\zeta}{t_0}\right)^{-\frac{7}{3}}\frac{2t_0^{\frac{5}{3}}+3\zeta^{\frac{5}{3}}}{t_0^{\frac{5}{3}}-t_1^{\frac{5}{3}}}\int_{t_0}^{t_1}d\xi\;\left(\frac{\xi}{t_0}\right)^{\frac{2}{3}}F(\xi)
\nonumber\\
u^{r(1)}(\zeta)&=&\frac{9}{20}t_0^2\left(\frac{\zeta}{t_0}\right)^{-\frac{2}{3}}\frac{\partial^2
F}{\partial r^2}(\zeta)-\frac{3}{10}t_0\left(\frac{\zeta}{t_0}\right)^{-1}\frac{\partial
F}{\partial r}(\zeta)\\&+&\frac{1}{10}\left(\frac{\zeta}{t_0}\right)^{-3}\left(\frac{t_1}{t_0}\right)^{\frac{5}{3}}\frac{4t_0^{\frac{5}{3}}+\zeta^{\frac{5}{3}}}{t_0^{\frac{5}{3}}-t_1^{\frac{5}{3}}}F(t_1)-\frac{1}{10}\left(\frac{\zeta}{t_0}\right)^{-3}\frac{4t_1^{\frac{5}{3}}+\zeta^{\frac{5}{3}}}{t_0^{\frac{5}{3}}-t_1^{\frac{5}{3}}}F(t_0)\nonumber\\&+&\frac{4}{5}\frac{1}{t_0}\left(\frac{\zeta}{t_0}\right)^{-3}\int_{t_0}^{\zeta}d\xi\;\left(\frac{\xi}{t_0}\right)^{\frac{2}{3}}F(\xi)-\frac{1}{5}\frac{1}{t_0}\left(\frac{\zeta}{t_0}\right)^{-3}\frac{4t_0^{\frac{5}{3}}+\zeta^{\frac{5}{3}}}{t_0^{\frac{5}{3}}-t_1^{\frac{5}{3}}}\int_{t_0}^{t_1}d\xi\;\left(\frac{\xi}{t_0}\right)^{\frac{2}{3}}F(\xi)\nonumber\\
u^{\vartheta(1)}(\zeta)&=&-\frac{1}{r^{(0)2}}\left(\frac{\zeta}{t_0}\right)^{-\frac{4}{3}}\int_{t_0}^{\zeta}d\xi\;\left[\frac{9}{20}\frac{t_0^2}{r^{(0)2}}\left(\frac{\partial
  F}{\partial
  \vartheta}-r^{(0)}\frac{\partial^2
  F}{\partial r\partial
  \vartheta}+\frac{r^{(0)2}}{2}\frac{\partial^3
  F}{\partial^2 r\partial
  \vartheta}\right)\right.\nonumber\\
&+&\left.\frac{3}{5}\frac{t_0}{r^{(0)}}\left(\frac{\xi}{t_0}\right)^{-\frac{1}{3}}\left(\frac{\partial
  F}{\partial
  \vartheta}-r^{(0)}\frac{\partial^2
  F}{\partial r\partial
  \vartheta}\right)-\frac{1}{2}\left(\frac{\xi}{t_0}\right)^{-\frac{2}{3}}\frac{\partial
  F}{\partial
  \vartheta}\right]\\
u^{\phi(1)}(\zeta)&=&-\frac{1}{r^{(0)2}\sin^2\vartheta}\left(\frac{\zeta}{t_0}\right)^{-\frac{4}{3}}\int_{t_0}^{\zeta}d\xi\;\left[\frac{9}{20}\frac{t_0^2}{r^{(0)2}}\left(\frac{\partial
  F}{\partial
  \phi}-r^{(0)}\frac{\partial^2
  F}{\partial r\partial
  \phi}+\frac{r^{(0)2}}{2}\frac{\partial^3
  F}{\partial^2 r\partial
  \phi}\right)\right.\nonumber\\
&+&\left.\frac{3}{5}\frac{t_0}{r^{(0)}}\left(\frac{\xi}{t_0}\right)^{-\frac{1}{3}}\left(\frac{\partial
  F}{\partial
  \phi}-r^{(0)}\frac{\partial^2
  F}{\partial r\partial
  \phi}\right)-\frac{1}{2}\left(\frac{\xi}{t_0}\right)^{-\frac{2}{3}}\frac{\partial
  F}{\partial
  \phi}\right]\\
w_{11}(\zeta)&=&-w_{22}(\zeta)\\
&=&\frac{\left(1-\left(\frac{t_1}{\zeta}\right)^{\frac{1}{3}}\right)^{-2}}{30\;\zeta^2\sin^2\vartheta}\int_{t_1}^{\zeta}d\xi\;\left(\frac{1-\left(\frac{t_1}{\xi}\right)^{\frac{1}{3}}}{1-\left(\frac{\xi}{t_0}\right)^{\frac{1}{3}}}\right)^2\left(\frac{\partial^2
F}{\partial \phi^2}-\sin^2\vartheta\frac{\partial^2
F}{\partial \vartheta^2}+\sin\vartheta\cos\vartheta\frac{\partial
F}{\partial \vartheta}\right)\nonumber\\
w_{12}(\zeta)&=&w_{21}(\zeta)\\
&=&-\frac{\left(1-\left(\frac{t_1}{\zeta}\right)^{\frac{1}{3}}\right)^{-2}}{15\;\zeta^2\sin^2\vartheta}\int_{t_1}^{\zeta}d\xi\;\left(\frac{1-\left(\frac{t_1}{\xi}\right)^{\frac{1}{3}}}{1-\left(\frac{\xi}{t_0}\right)^{\frac{1}{3}}}\right)^2\left(\cos\vartheta\frac{\partial
F}{\partial \phi}-\sin\vartheta\frac{\partial^2
F}{\partial \vartheta\partial \phi}\right)\nonumber
\end{eqnarray}
The first order corrections to the Hubble diagram are given by
\begin{eqnarray}
z^{(1)}&=&\frac{3}{10}t_0\left(\frac{t_1}{t_0}\right)^{-\frac{2}{3}}\left[\left(\frac{t_1}{t_0}\right)^{\frac{1}{3}}\frac{\partial
F}{\partial r}(t_1)-\frac{\partial
F}{\partial
  r}(t_0)\right]\nonumber\\
&+&\frac{1}{10}\left(\frac{t_1}{t_0}\right)^{-\frac{2}{3}}\left(F(t_1)-F(t_0)\right)
\label{sw}
\end{eqnarray}
\begin{eqnarray}
&&\ln I-\ln I^{(0)}(z)=\left(\frac{t_1}{t_0}\right)^{\frac{2}{3}}\frac{z^{(1)}}{\left(\frac{t_1}{t_0}\right)^{-\frac{1}{3}}-1}-\frac{A^{(1)}}{A^{(0)}}\nonumber\\
&=&-\frac{3}{10}t_0\left(\frac{t_1}{t_0}\right)^{\frac{1}{3}}\frac{t_0^{\frac{1}{3}}\frac{\partial
F}{\partial
  r}(t_0)-t_1^{\frac{1}{3}}\frac{\partial
F}{\partial r}(t_1)}{t_0^{\frac{1}{3}}-t_1^{\frac{1}{3}}}\nonumber\\
&+&\frac{1}{10}\left(\frac{t_0^{-\frac{1}{3}}}{t_1^{-\frac{1}{3}}-t_0^{-\frac{1}{3}}}
-2\frac{\left(t_1^{-\frac{1}{3}}-t_0^{-\frac{1}{3}}\right)\left(t_1^{-\frac{2}{3}}-t_0^{-\frac{2}{3}}\right)\left(2t_1^{-\frac{2}{3}}+t_1^{-\frac{1}{3}}t_0^{-\frac{1}{3}}+2t_0^{-\frac{2}{3}}\right)}{t_1^{-\frac{5}{3}}-t_0^{-\frac{5}{3}}}\right)\left(F(t_1)-F(t_0)\right)\nonumber\\&-&2\frac{t_1^{-\frac{5}{3}}t_0^{-\frac{5}{3}}}{t_1^{-\frac{5}{3}}-t_0^{-\frac{5}{3}}}\int_{t_1}^{t_0}d\xi\;\xi^{\frac{2}{3}}F(\xi)+\frac{2}{5}\frac{1}{t_0^{\frac{1}{3}}-t_1^{\frac{1}{3}}}\int_{t_1}^{t_0}d\xi\;\xi^{-\frac{2}{3}}F(\xi)\nonumber\\&-&\frac{9}{5}\frac{t_0^{\frac{4}{3}}}{t_0^{\frac{1}{3}}-t_1^{\frac{1}{3}}}\int_{t_1}^{t_0}d\xi\;\triangle
F(\xi) \left(-1+(t_0^{\frac{1}{3}}+t_1^{\frac{1}{3}})\xi^{-\frac{1}{3}}-t_0^{\frac{1}{3}}t_1^{\frac{1}{3}}\xi^{-\frac{2}{3}}\right)
\label{e4.h1}
\end{eqnarray}
or, in terms of the luminosity distance,
\begin{eqnarray}
&&d_L-d_L^{(0)}(z)=\frac{1}{2}d_L^{(0)}(z^{(0)})\left[\frac{A^{(1)}}{A^{(0)}}-\frac{z^{(1)}}{\left(\frac{t_1}{t_0}\right)^{-\frac{1}{3}}-1}\right]\nonumber\\
&=&\frac{3}{2}t_0\left(1-\left(\frac{t_1}{t_0}\right)^{\frac{1}{3}}\right)\left(\frac{t_1}{t_0}\right)^{-\frac{2}{3}}\left[
\frac{3}{10}t_0\left(\frac{t_1}{t_0}\right)^{-\frac{1}{3}}\frac{t_0^{\frac{1}{3}}\frac{\partial
F}{\partial
  r}(t_0)-t_1^{\frac{1}{3}}\frac{\partial
F}{\partial r}(t_1)}{t_0^{\frac{1}{3}}-t_1^{\frac{1}{3}}}\right.\nonumber\\
&-&\left.\frac{1}{10}\left(\frac{t_1^{-\frac{2}{3}}t_0^{\frac{1}{3}}}{t_1^{-\frac{1}{3}}-t_0^{-\frac{1}{3}}}
-2\frac{\left(t_1^{-\frac{1}{3}}-t_0^{-\frac{1}{3}}\right)\left(t_1^{-\frac{2}{3}}-t_0^{-\frac{2}{3}}\right)\left(2t_1^{-\frac{2}{3}}+t_1^{-\frac{1}{3}}t_0^{-\frac{1}{3}}+2t_0^{-\frac{2}{3}}\right)}{t_1^{-\frac{5}{3}}-t_0^{-\frac{5}{3}}}\right)\left(F(t_1)-F(t_0)\right)\right.\nonumber\\&+&\left.2\frac{t_1^{-\frac{5}{3}}t_0^{-\frac{5}{3}}}{t_1^{-\frac{5}{3}}-t_0^{-\frac{5}{3}}}\int_{t_1}^{t_0}d\xi\;\xi^{\frac{2}{3}}F(\xi)-\frac{2}{5}\frac{1}{t_0^{\frac{1}{3}}-t_1^{\frac{1}{3}}}\int_{t_1}^{t_0}d\xi\;\xi^{-\frac{2}{3}}F(\xi)\right.\nonumber\\&+&\left.\frac{9}{5}\frac{t_0^{\frac{4}{3}}}{t_0^{\frac{1}{3}}-t_1^{\frac{1}{3}}}\int_{t_1}^{t_0}d\xi\;\triangle
F(\xi)
\left(-1+(t_0^{\frac{1}{3}}+t_1^{\frac{1}{3}})\xi^{-\frac{1}{3}}-t_0^{\frac{1}{3}}t_1^{\frac{1}{3}}\xi^{-\frac{2}{3}}\right)\right]\nonumber\\
&\approx&\frac{1}{2}d_L^{(0)}(z)\left[\frac{3}{10}t_0\sqrt{1+z}\frac{\sqrt{1+z}\frac{\partial
F}{\partial
  r}(t_0)-\frac{\partial
F}{\partial r}(t_1)}{\sqrt{1+z}-1}\right.\nonumber\\
&-&\left.\frac{1}{10}\left(\frac{1+z}{\sqrt{1+z}-1}
-2\frac{z\left(2+\sqrt{1+z}+2(1+z)\right)}{1+\sqrt{1+z}+(1+z)+\left(\sqrt{1+z}\right)^3+(1+z)^2}\right)\left(F(t_1)-F(t_0)\right)\right.\nonumber\\&+&\left.2t_0^{-\frac{5}{3}}\frac{\left(1+z\right)^{5/2}}{\left(1+z\right)^{5/2}-1}\int_{t_0(1+z)^{-3/2}}^{t_0}d\xi\;\xi^{\frac{2}{3}}F(\xi)-\frac{2}{5}t_0^{-\frac{1}{3}}\frac{\sqrt{1+z}}{\sqrt{1+z}-1}\int_{t_0(1+z)^{-3/2}}^{t_0}d\xi\;\xi^{-\frac{2}{3}}F(\xi)\right.\nonumber\\&+&\left.\frac{9}{5}t_0\frac{\sqrt{1+z}}{\sqrt{1+z}-1}\int_{t_0(1+z)^{-3/2}}^{t_0}d\xi\;\triangle
F(\xi)
\left(-1+t_0^{\frac{1}{3}}\left(1+\frac{1}{\sqrt{1+z}}\right)\xi^{-\frac{1}{3}}-\frac{t_0^{\frac{2}{3}}}{\sqrt{1+z}}\xi^{-\frac{2}{3}}\right)\right]
\label{e4.h2}
\end{eqnarray}
Note that Eq.(\ref{sw}) is just the Sachs-Wolfe effect\cite{Sachs-Wolfe}.
\section{Second order}
First of all, we need the second order correction to the path and the four velocity $u^i$. 
Again, we use both Eq.(\ref{e1.2}) and Eq.(\ref{e1.3}). We get from Eq.(\ref{e1.2})
\begin{eqnarray}
\frac{2}{a}\frac{dt^{(2)}}{d\lambda}+2\frac{dr^{(2)}}{d\lambda}-\frac{2\dot a}{a^3}t^{(2)}=-\frac{g^{(2)}_{rr}}{a^4}
-2g^{(1)}_{ij}u^{i(1)}u^{j(0)}-g^{(0)}_{ij}u^{i(1)}u^{j(1)}\nonumber\\
-2\frac{\partial g^{(0)}_{ij}}{\partial x^l}x^{l(1)}u^{i(1)}u^{j(0)}
-\frac{1}{a^4}\frac{\partial g^{(1)}_{rr}}{\partial x^l}x^{l(1)}
-\frac{1}{2a^4}\frac{\partial^2 g^{(0)}_{rr}}{\partial x^l\partial x^m}x^{l(1)}x^{m(1)}\;.
\label{e5.1}
\end{eqnarray}
Similarly, Eq.(\ref{e1.3}) leads to
\begin{eqnarray}
\frac{d^2t^{(2)}}{d\lambda^2}-2\frac{\dot a}{a}\frac{dr^{(2)}}{d\lambda}
+\left(\frac{\ddot a}{a^3}+\frac{\dot
  a^2}{a^4}\right)t^{(2)}=-\Gamma^{t(2)}_{jk}u^{j(0)}u^{k(0)}\nonumber\\
-2\Gamma^{t(1)}_{jk}u^{j(1)}u^{k(0)}-\Gamma^{t(0)}_{jk}u^{j(1)}u^{k(1)}
-2\frac{\partial \Gamma^{t(0)}_{jk}}{\partial x^l}x^{l(1)}u^{j(1)}u^{k(0)}\nonumber\\
-\frac{\partial \Gamma^{t(1)}_{jk}}{\partial x^l}x^{l(1)}u^{j(0)}u^{k(0)}
-\frac{1}{2}\frac{\partial^2 \Gamma^{t(0)}_{jk}}{\partial x^l\partial x^m}x^{l(1)}x^{m(1)}u^{j(0)}u^{k(0)}
\label{e5.2a}\\
\frac{d^2r^{(2)}}{d\lambda^2}-2\frac{\dot a}{a^3}\frac{dt^{(2)}}{d\lambda}+2\frac{\dot a}{a^2}\frac{dr^{(2)}}{d\lambda}
-2\left(\frac{\ddot a}{a^4}-\frac{\dot
  a^2}{a^5}\right)t^{(2)}=-\Gamma^{r(2)}_{jk}u^{j(0)}u^{k(0)}\nonumber\\
-2\Gamma^{r(1)}_{jk}u^{j(1)}u^{k(0)}-\Gamma^{r(0)}_{jk}u^{j(1)}u^{k(1)}
-2\frac{\partial \Gamma^{r(0)}_{jk}}{\partial x^l}x^{l(1)}u^{j(1)}u^{k(0)}\nonumber\\
-\frac{\partial \Gamma^{r(1)}_{jk}}{\partial x^l}x^{l(1)}u^{j(0)}u^{k(0)}
-\frac{1}{2}\frac{\partial^2 \Gamma^{r(0)}_{jk}}{\partial x^l\partial x^m}x^{l(1)}x^{m(1)}u^{j(0)}u^{k(0)}\label{e5.2b}\\
\frac{d^2\vartheta^{(2)}}{d\lambda^2}+2\frac{\dot{(ar^{(0)})}}{a^2r^{(0)}}\frac{d\vartheta^{(2)}}{d\lambda}
=-\Gamma^{\vartheta(2)}_{jk}u^{j(0)}u^{k(0)}\nonumber\\
-2\Gamma^{\vartheta(1)}_{jk}u^{j(1)}u^{k(0)}-\Gamma^{\vartheta(0)}_{jk}u^{j(1)}u^{k(1)}
-2\frac{\partial \Gamma^{\vartheta(0)}_{jk}}{\partial x^l}x^{l(1)}u^{j(1)}u^{k(0)}\nonumber\\
-\frac{\partial \Gamma^{\vartheta(1)}_{jk}}{\partial x^l}x^{l(1)}u^{j(0)}u^{k(0)}
-\frac{1}{2}\frac{\partial^2 \Gamma^{\vartheta(0)}_{jk}}{\partial x^l\partial x^m}x^{l(1)}x^{m(1)}u^{j(0)}u^{k(0)}\label{e5.2c}\\
\frac{d^2\phi^{(2)}}{d\lambda^2}+2\frac{\dot{(ar^{(0)})}}{a^2r^{(0)}}\frac{d\phi^{(2)}}{d\lambda}
=-\Gamma^{\phi(2)}_{jk}u^{j(0)}u^{k(0)}\nonumber\\
-2\Gamma^{\phi(1)}_{jk}u^{j(1)}u^{k(0)}-\Gamma^{\phi(0)}_{jk}u^{j(1)}u^{k(1)}
-2\frac{\partial \Gamma^{\phi(0)}_{jk}}{\partial x^l}x^{l(1)}u^{j(1)}u^{k(0)}\nonumber\\
-\frac{\partial \Gamma^{\phi(1)}_{jk}}{\partial x^l}x^{l(1)}u^{j(0)}u^{k(0)}
-\frac{1}{2}\frac{\partial^2 \Gamma^{\phi(0)}_{jk}}{\partial x^l\partial x^m}x^{l(1)}x^{m(1)}u^{j(0)}u^{k(0)}\label{e5.2d}
\end{eqnarray} 
Since the left hand sides are of the same form as in the case of the first corrections, while the right hand sides are 
already known, these equations can be solved by the same technique as Eqs.(\ref{e4.1a})-(\ref{e4.2d}). Once the path and the four velocity are known to second order, one can determine the second correction to the expansion rate as well.
The Sachs optical equation implies (cf. Eq.(\ref{e4.14}))
\begin{eqnarray}
\frac{1}{a}\frac{d\Theta^{(2)}}{d\zeta}+2\frac{\dot{(a d)}}{a^2 d}\Theta^{(2)}=\chi^{(2)}(\zeta)
\label{e5.3}
\end{eqnarray} 
with
\begin{eqnarray}
\chi^{(2)}=-\frac{1}{2}R_{jk}^{(2)}u^{j(0)}u^{k(0)}-R_{jk}^{(1)}u^{j(1)}u^{k(0)}
-\frac{1}{2}\frac{\partial R_{jk}^{(1)}}{\partial x^l}x^{l(1)}u^{j(0)}u^{k(0)}\nonumber\\
-R_{jk}^{(0)}u^{j(2)}u^{k(0)}-\frac{1}{2}R_{jk}^{(0)}u^{j(1)}u^{k(1)}
-\frac{\partial R_{jk}^{(0)}}{\partial x^l}x^{l(1)}u^{j(1)}u^{k(0)}\nonumber\\
-\frac{1}{2}\frac{\partial R_{jk}^{(0)}}{\partial x^l}x^{l(2)}u^{j(0)}u^{k(0)}
-\frac{1}{4}\frac{\partial^2 R_{jk}^{(0)}}{\partial x^l\partial x^m}x^{l(1)}x^{m(1)}u^{j(0)}u^{k(0)}\nonumber\\
-\frac{1}{2}w^{(1)2}-\Theta^{(1)2}\;.
\label{e5.4}
\end{eqnarray} 
As before, the solution of Eq.(\ref{e5.3}) is
\begin{eqnarray}
\Theta^{(2)}=\frac{1}{a^2 d^{2}}\int_{t_1}^{\zeta}d\xi\;a^3 d^{2}\chi^{(2)}(\xi)\;.
\label{e5.5}
\end{eqnarray}
For the second order correction of the beam area we get the equation
\begin{eqnarray}
\frac{1}{a}\frac{dA^{(2)}}{d\zeta}-2\frac{\dot{(ad)}}{a^2 d}A^{(2)}=2\Theta^{(2)}\Omega a^2d^{2}
+2\Theta^{(1)}A^{(1)}\;.
\label{e5.6}
\end{eqnarray}
Its solution is
\begin{eqnarray}
A^{(2)}=2a^2 d^{2}
\int_{t_1}^{\zeta}d\xi\;\left(a \Omega \Theta^{(2)}+\frac{1}{ad^{2}}\Theta^{(1)}A^{(1)}\right)\;.
\label{e5.7}
\end{eqnarray}
\newpage
\section{How does the Hubble diagram change?}
We consider now the question how the Hubble diagram changes due to the
perturbation. As the Hubble diagram is given in a parametric form, we calculate
the difference
\begin{eqnarray}
\ln{I}-\ln{I^{(0)}(z)}\;,\label{e6.1}
\end{eqnarray}  
i.e., we compare the perturbed intensity with that of the nonperturbed one,
taking the latter at the perturbed redshift value. Namely,
\begin{eqnarray}
z=z^{(0)}+z^{(1)}+z^{(2)}\;\label{e6.2}
\end{eqnarray} 
where
\begin{eqnarray}
z^{(0)}=\left(\frac{t_0}{t_1}\right)^{\frac{2}{3}}-1\;,\label{e6.3a}\\
z^{(1)}=a(t_0)\left[u^{t(1)}(t_1)-(1+z^{(0)})u^{t(1)}(t_0)\right]\;,\label{e6.3b}\\
z^{(2)}=a(t_0)\left[u^{t(2)}(t_1)-(1+z^{(0)})u^{t(2)}(t_0)\right]\nonumber\\
-a^4(t_0)u^{t(1)}(t_0)\left[u^{t(1)}(t_1)-(1+z^{(0)})u^{t(1)}(t_0)\right]\;.\label{e6.3c}
\end{eqnarray}
In this way we obtain
\begin{eqnarray}
\ln{I}-\ln{I^{(0)}(z)}=
\left[\frac{z^{(1)}}{(1+z^{(0)})\left(\sqrt{1+z^{(0)}}-1\right)}
-\frac{(1+z^{(0)})A^{(1)}}{9t_0^2\left(\sqrt{1+z^{(0)}}-1\right)^2}
\right]
\nonumber\\
+\left[\frac{z^{(2)}}{(1+z^{(0)})\left(\sqrt{1+z^{(0)}}-1\right)}
-\frac{(1+z^{(0)})A^{(2)}}{9t_0^2\left(\sqrt{1+z^{(0)}}-1\right)^2}\right.\nonumber\\\left.
+\frac{\left(2-3\sqrt{1+z^{(0)}}\right)z^{(1)2}}{4(1+z^{(0)})^2\left(\sqrt{1+z^{(0)}}-1\right)^2}
+\frac{(1+z^{(0)})^2A^{(1)2}}{162t_0^4\left(\sqrt{1+z^{(0)}}-1\right)^4}
\right]
\label{e6.4}
\end{eqnarray}  

\section{Acknowledgement}
The present work was supported by the OTKA grant NI 68228.

\appendix
\section{Calculating $A^{(1)}$}
First we make the expression of $\chi^{(1)}$ more explicite.
According to Eqs.(\ref{e4.13}), (\ref{e4.14})
\begin{eqnarray}
\chi^{(1)}(\zeta)=-\frac{1}{2}R_{jk}^{(1)}u^{j(0)}u^{k(0)}
-R_{jk}^{(0)}u^{j(1)}u^{k(0)}-\frac{1}{2}\frac{\partial R_{jk}^{(0)}}{\partial x^l}x^{l(1)}u^{j(0)}u^{k(0)}
\label{eA.1}
\end{eqnarray}
Inserting the zeroth order expressions, we have
\begin{eqnarray}
\chi^{(1)}(\zeta)=-\frac{1}{2a^2}R_{tt}^{(1)} +\frac{1}{a^3}R_{tr}^{(1)} -\frac{1}{2a^4}R_{rr}^{(1)} 
+3\frac{\ddot a}{a^2}u^{t(1)}+\frac{a \ddot a+2\dot a^2}{a^2}u^{r(1)}+\frac{a \stackrel{\textrm{...}}{a}-4\dot a \ddot a}{a^4}t^{(1)}
\label{eA.2}
\end{eqnarray}
Here 
\begin{eqnarray}
R_{tt}^{(1)}=-\frac{21}{10}\triangle B-30\frac{a_0}{a}B
\label{eA.3}
\end{eqnarray}
\begin{eqnarray}
R_{tr}^{(1)}=-2(a_0 a)^{1/2}\frac{\partial B}{\partial r}
\label{eA.4}
\end{eqnarray}
\begin{eqnarray}
R_{rr}^{(1)}=\frac{9}{10}\frac{a_0}{a}\zeta^2\triangle B
+\frac{126}{5}\frac{a_0}{a}\zeta^2\frac{\partial^2 B}{\partial r^2}
+288\left(\frac{a_0}{a}\right)^2\zeta^2B
\label{eA.5}
\end{eqnarray}
Combining Eqs.(\ref{eA.2})-(\ref{eA.5}), (\ref{e4.8})-(\ref{e4.12}) we get
\begin{eqnarray}
\chi^{(1)}(\zeta)&=&-\frac{1}{2a^2}R_{tt}^{(1)} +\frac{1}{a^3}R_{tr}^{(1)} -\frac{1}{2a^4}R_{rr}^{(1)}\nonumber\\ 
&&-\frac{4}{3}\frac{1}{a^2\zeta^{11/3}}\int_{t_0}^{\zeta}d\xi\;\xi^{\frac{5}{3}}f(\xi)
-\frac{1}{3}\frac{g^{(1)}_{rr}}{a^4\zeta^{2}}
+\frac{20}{9}\frac{\beta}{a^2\zeta^{11/3}}
\label{eA.6}
\end{eqnarray}

\end{document}